\documentclass[11pt,twoside]{article}
\usepackage{asp2010}
\usepackage{graphicx}
\usepackage{natbib}

\resetcounters

\begin{document}

\title{	Detailed abundances in extremely metal poor dwarf stars extracted from SDSS}
\author{L. Sbordone,$^{1,2}$ P. Bonifacio$^2$, E. Caffau,$^{1,2}$, and H.-G. Ludwig$^{1,2}$
\affil{$^1$Zentrum f\"ur Astronomie der Universit\"at Heidelberg, Landessternwarte,
K\"onigstuhl 12, 69117 Heidelberg, Germany}
\affil{$^2$GEPI, Observatoire de Paris, CNRS, Univ. Paris Diderot, Place
Jules Janssen, 92190
Meudon, France}}

\begin{abstract}
We report on the result of an ongoing campaign to determine chemical abundances in extremely metal poor (EMP) turn-off (TO) stars selected from the Sloan Digital Sky Survey (SDSS) low resolution spectra. This contribution focuses principally on the largest part of the sample (18 stars out of 29), observed with UVES@VLT and analyzed by means of the automatic abundance analysis code MyGIsFOS to derive atmosphere parameters and detailed compositions. The most significant findings include i) the detection of a C-rich, strongly Mg-enhanced star ([Mg/Fe]=1.45); ii) a group of Mn-rich stars ([Mn/Fe]$>$-0.4); iii) a group of Ni-rich stars ([Ni/Fe]$>$0.2). Li is measured in twelve stars, while for three upper limits are derived.
\end{abstract}

\section{Introduction}
Since many years we are conducting a campaign to select EMP candidates from the Sloan Digital Sky Survey \citep[SDSS][]{york00}. Candidate TO stars are selected among SDSS spectroscopic targets on the basis of $(g-z)_0$ color, and their SDSS low resolution spectra are screened by means of an automatic procedure to derive a metallicity estimate \citep{ludwig08}. A number of candidates have been so far observed by means of the X-Shooter \citep{dodorico06} and UVES \citep{dekker00} spectrographs at VLT, and partial results have appeared in the literature. \cite{behara10} presented 3 carbon enhanced metal poor stars observed with UVES, and \cite{bonifacio11} and \citet{caffau11a} described 7 stars observed with X-Shooter. The most striking result so far has been the discovery of SDSS J1029151+172927 \citep{caffau11b}, an ultra metal poor (UMP)\footnote{EMP / UMP and similar designations were initially established as being based on [Fe/H] only \citep{beers05}. As such, they are misleading due to the extreme CNO enrichment often seen in stars of extremely low iron content. However, we keep using them here in the usual meaning for consistency.}
TO star ([Fe/H]$_\mathrm{3D}$=-5.0) with an abundance pattern typical of Halo stars for all measured elements. The CNO elements are not measured, but strong enhancements can be excluded ([C/Fe]$<$0.7, [N/Fe]$<$0.2).
 This characteristic, unique among the 4 UMP stars known to date, sets the current record of the star with the lowest metal content (Z$<$4.5 10$^{-5}$Z$_{\sun}$). The discovery of this object challenges the \citet{bromm03} theory of low metallicity star formation, which implies that a higher C and O abundance would be necessary to form a star of such low mass. On the other hand, it favors theories predicting a lower critical metallicity for the formation of low mass stars \citep[e.g.][]{schneider12} or even the absence of a critical metallicity \citep[][]{greif11}.

In this contribution we will focus on the bulk of the sample of stars observed with UVES@VLT (18 stars) about which a more extensive paper is in preparation \citep{bonifacio12}.

\articlefigurefour{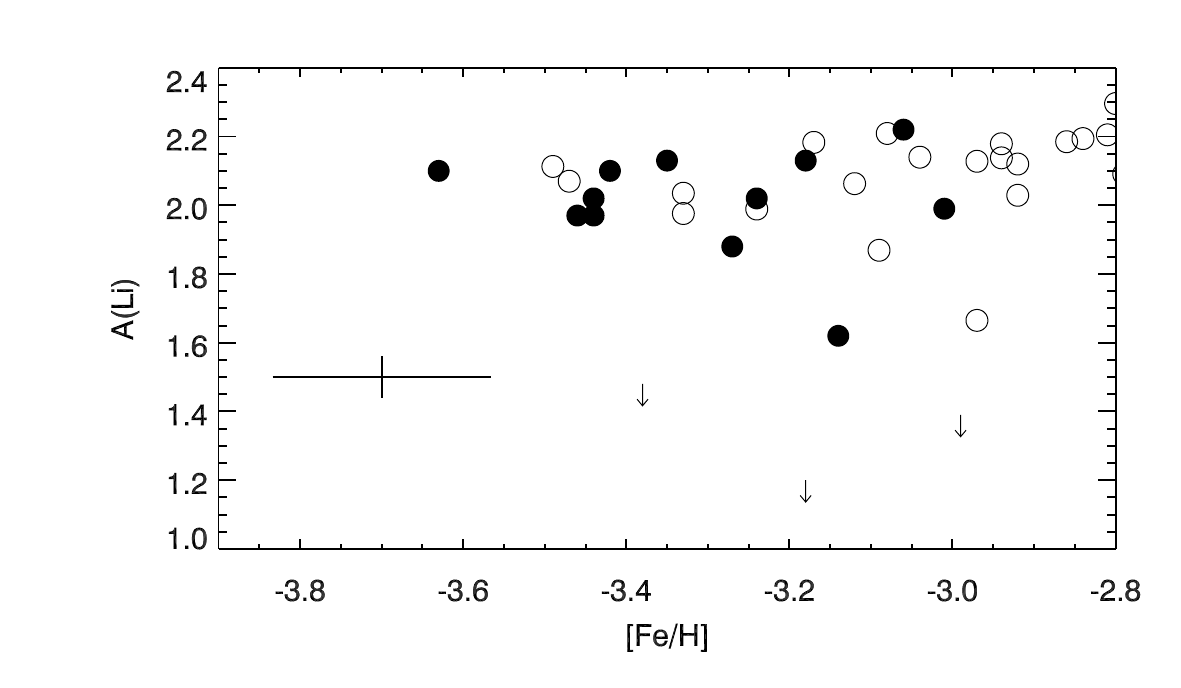}{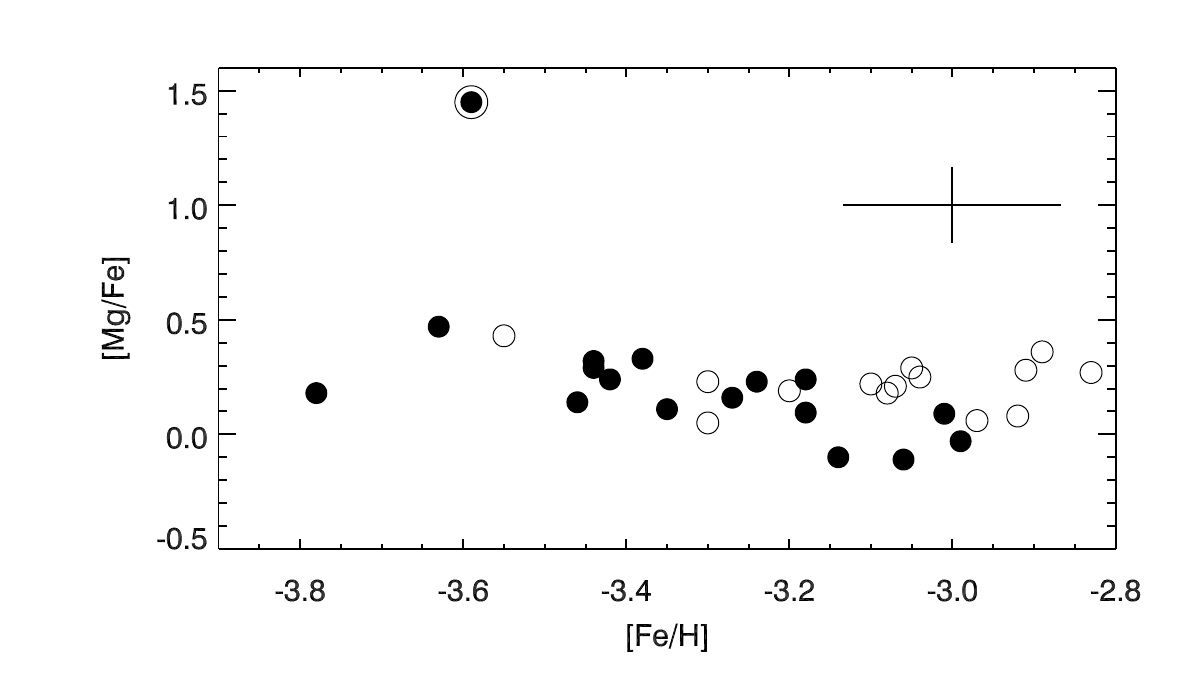}{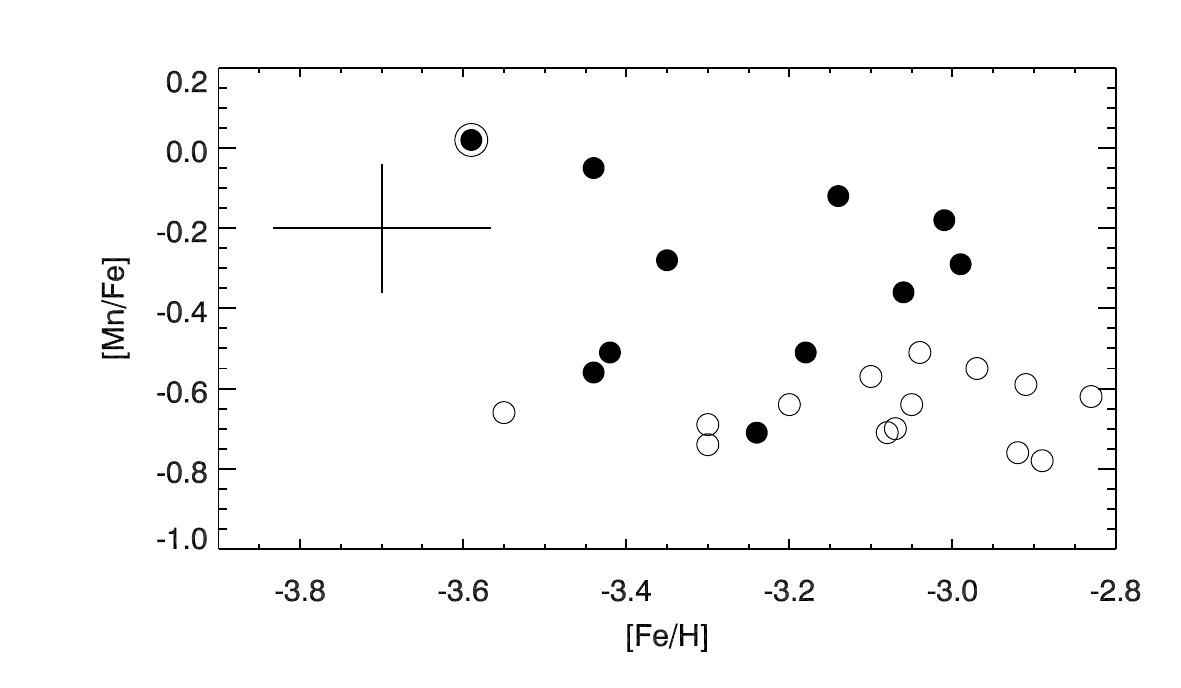}{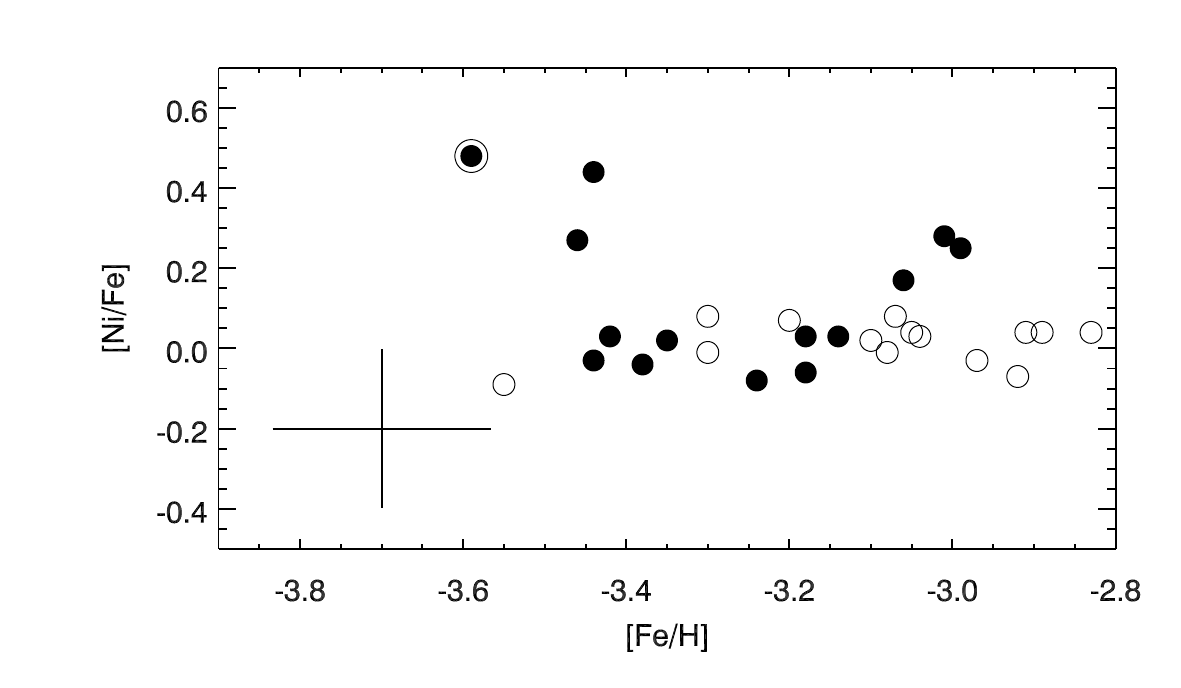}
{theplots}{Abundance determinations for Li, Mg, Mn and Ni plotted against [Fe/H] for the stars in the UVES SDSS sample (filled circles, the encircled point indicates the preliminary results for SDSS J134922+140736). Open circles represent literature data. Upper left panel: A(Li) measurements or upper limits, against \cite{sbordone10} results (3D temperature scale parameters and A(Li)$_\mathrm{3D,NLTE}$); other panels [Mg/Fe], [Mn/Fe], [Ni/Fe] against \cite{bonifacio09} data for EMP TO stars. Typical error bars superimposed.}

\section{Observations and analysis}
The observations were performed with UVES set in 380+580nm dichroic mode, 1.4'' slit and 2$\times$2 pixel binning. Reduction was carried out by means of the standard ESO UVES pipeline. The analysis was performed with the {\tt MyGIsFOS} code \citep{sbordone10b,sbordone12a}. This is an evolution of the code of \citet{bonifacio03}, with a number of improvements, most notably a loop to determine surface gravity from the iron ionization equilibrium. The grid of synthetic spectra was computed from a grid of ATLAS 12 \citep{kurucz05,castelli05,sbordone04,sbordone05} model atmospheres. 
The effective temperature was considered a prior and was derived from the $(g-z)_0$ color using the calibration presented in \citet{ludwig08}. Surface gravity was determined
from the iron ionization equilibrium. If no \ion{Fe}{ii} lines were retained in the analysis, the surface gravity was held fixed at $\log{g}$ = 4.0.

One star (SDSS J134922+140736) turned out to be C-rich (qualitative estimate, from G-band strength), and the {\tt MyGIsFOS} analysis determines an extreme Mg enrichment ([Mg/Fe]=1.45). The whole analysis technique employed was not designed to handle C-rich stars. Moreover, {\tt MyGIsFOS} is potentially prone to systematic deviations when measuring abundances strongly deviating from the ones assumed in the grid (such as is the case of Mg in this star). As a consequence, until an better adapted analysis is performed for this star \citep{sbordone12b}, results for SDSS J134922+140736 are to be considered preliminary.

Lithium abundance was measured using the same technique employed in \citet{sbordone10}: the 670.7nm doublet equivalent width was estimated by fitting a grid of synthetic profiles, and the \citet{sbordone10} fitting function was applied to derive a 3D, NLTE Li abundance measurement.

One additional star was observed (SDSS J153110+095255) but resulted to be a double-lined spectroscopic binary: it is thus excluded from the analysis.

\section{Results and discussion}
In fig. 1 we show some of the most interesting results of the present analysis. A more through analysis is to appear in \citet{bonifacio12} and \citet{sbordone12b}. 

{\bf Lithium}: in \citet{sbordone10} we identified a progressive disruption of the Spite plateau, taking place below [Fe/H]$\sim$-2.8. We referred to this observed feature as a ``meltdown'': an increasing number of stars show some variable degree of Li depletion. The new observations reinforce this picture. \citet{melendez10} suggested that convective Li depletion should be expected in stars as warm as $\sim$6100 K in the EMP range. This is partially compatible with the results in \citet{sbordone10}, where every star below 6000 K is Li depleted. However, the new sample contains one object (SDSS J090733+024608) whose Li abundance places it on the Spite plateau despite a T$_{\rm eff}$=5934 K, and both the \citet{sbordone10} sample and the new one contain a number of Li-poor objects of relatively high temperature. Moreover, SDSS J1029151+172927, otherwise a very inconspicuous star, shares the same deep Li depletion as HE 1327--2326 \citep[][{[Fe/H]=-5.96,[C/Fe]=+3.78}]{frebel08}. This leads us to believe that a fundamental understanding of the low metallicity  Li behavior is still eluding us. 

{\bf Mg, and the chemically peculiar SDSS J134922+140736}: the new sample pre-sents for the most part the usual pattern of $\alpha$-enhancement expected in the metal poor Halo. However, three $\alpha$-poor stars are found, similar, but at lower metallicity, to the two stars described in \citet{bonifacio11}. The extreme Mg overabundance is the most striking anomaly of SDSS J134922+140736, but it is interesting to notice that the star appears anomalous in Ni and Mn too, as well as C-rich. Two similarly Mg-rich stars have been detected also by the 0Z survey \citep[][and J. Cohen's contribution at this conference]{cohen11}, and a possibly similar star was detected by the ``First Stars'' ESO Large Program \citep{depagne02}. Li is not detected due to the low S/N in the red region of the spectrum, but if measured should provide a diagnostic of the amount of high-temperature processed gas that polluted the star.

{\bf Mn and Ni rich populations}: while the detection of a Ni rich subsample is not highly significant due to the associated error, a Mn overabundant subsample is evident also by comparing the Mn line strengths in stars with otherwise very close parameters and metallicities. Both anomalies are reported also in J. Cohen's contribution to this conference, as detected by the 0Z survey. There is a hint of a correlation between Ni and Mn enrichment: of the seven stars with [Mn/Fe]$>$-0.4, four have [Ni/Fe]$>$0.2. The size of the errors and the limited sample, however, prevent us from deriving more stringent conclusions.

\acknowledgements PB acknowledges support from the Programme National
de Physique Stellaire (PNPS) and the Programme National
de Cosmologie et Galaxies (PNCG) of the Institut National de Sciences
de l'Universe of CNRS.

\end{document}